\documentclass[conference]{IEEEtran}

\IEEEoverridecommandlockouts

\usepackage{cite}
\usepackage{bm}
\usepackage{amsmath}
\usepackage{algorithmic}
\usepackage{graphicx}
\usepackage{textcomp}
\usepackage{xcolor}
\def\BibTeX{{\rm B\kern-.05em{\sc i\kern-.025em b}\kern-.08em
    T\kern-.1667em\lower.7ex\hbox{E}\kern-.125emX}}

\usepackage[utf8]{inputenc}
\usepackage[american]{babel}
\usepackage[hidelinks]{hyperref}
\usepackage[capitalize]{cleveref}
\usepackage{mathtools}
\usepackage{booktabs}
\usepackage{fancyhdr}
\usepackage[shortcuts]{extdash}
\usepackage[all]{foreign}
\usepackage{bm}
\usepackage{tikz}
\usetikzlibrary{bayesnet}
\usepackage{siunitx}
\usepackage{xfrac}

\usepackage{booktabs}
\usepackage{arydshln}
\usepackage{comment}
\usepackage{subcaption}

\usepackage[helvratio=0.92]{newtxtext}

\usepackage{fancyhdr}

\usepackage{glossaries}

\newacronym{ad}{AD}{Annotated Disjunction}
\newacronym{ap}{AP}{Access Point}
\newacronym{cfr}{CFR}{channel frequency response}
\newacronym{cnn}{CNN}{convolutional neural network}
\newacronym{csi}{CSI}{channel state information}
\newacronym{cv}{CV}{computer vision}
\newacronym{dl}{DL}{deep learning}
\newacronym{edl}{EDL}{Evidential Deep Learning}
\newacronym{har}{HAR}{human activity recognition}
\newacronym{kl}{\mbox{KL}}{Kullback–Leibler}
\newacronym{lan}{LAN}{local-area network}
\newacronym{lstm}{LSTM}{long short-term memory}
\newacronym{mimo}{MIMO}{multiple-input multiple-output}
\newacronym{mlp}{MLP}{multi-layer perceptron}
\newacronym{nad}{nAD}{Neural Annotated Disjunction}
\newacronym{nic}{NIC}{network interface card}
\newacronym{nn}{NN}{neural network}
\newacronym{ood}{OoD}{out-of-distribution}
\newacronym{ofdm}{OFDM}{orthogonal frequency-division multiplexing}
\newacronym{ofdma}{OFDMA}{orthogonal frequency-division multiple access}
\newacronym{phy}{PHY}{Physical Layer}
\newacronym{sdr}{SDR}{software-defined radio}
\newacronym{siso}{SISO}{single-input single-output}
\newacronym{sta}{STA}{station}
\newacronym{vae}{VAE}{Variational Auto-Encoder}
\newacronym{wlan}{WLAN}{wireless local-area network}
\newcommand*\wifi{\mbox{Wi-Fi}\xspace}

\newcommand{\limbangle}{\ensuremath{\alpha}\xspace}
\newcommand{\anglefeature}{\ensuremath{\delta}\xspace}

\newcommand{\vaef}{\mbox{VAE-F}\xspace}

\newcommand{\deepprobhar}{\mbox{\textsf{DeepProbHAR}}\xspace}

\newcommand{\nofusing}[1]{\mbox{No-Fused-$\mathrm{#1}$}\xspace}
\newcommand{\earlyfusing}{\mbox{Early-Fusing}\xspace}
\newcommand{\delayedfusing}{\mbox{Delayed-Fusing}\xspace}

\newcommand{\boldnofusing}[1]{\mbox{\textbf{No-Fused-}$\mathbf{#1}$}\xspace}
\newcommand{\boldearlyfusing}{{\bf \earlyfusing}}
\newcommand{\bolddelayedfusing}{{\bf \delayedfusing}}

\begin{document}

\title{Neuro-Symbolic Fusion of Wi-Fi Sensing Data for Passive Radar with Inter-Modal Knowledge Transfer}

\author{
    \IEEEauthorblockN{
        Marco Cominelli\IEEEauthorrefmark{1},
        Francesco Gringoli\IEEEauthorrefmark{2},
        Lance M. Kaplan\IEEEauthorrefmark{3}, 
        Mani B. Srivastava\IEEEauthorrefmark{4},\\
        Trevor Bihl\IEEEauthorrefmark{6},
        Erik P. Blasch\IEEEauthorrefmark{6},
        Nandini Iyer\IEEEauthorrefmark{6},
        and Federico Cerutti\IEEEauthorrefmark{2}
    }\\
    \IEEEauthorblockA{
        \IEEEauthorrefmark{1}%
            DEIB, 
            Politecnico di Milano, Italy. 
            marco.cominelli@polimi.it
    }
    \IEEEauthorblockA{
        \IEEEauthorrefmark{2}%
            DII, 
            University of Brescia, Italy. 
            \{francesco.gringoli, federico.cerutti\}@unibs.it
    }
    \IEEEauthorblockA{
        \IEEEauthorrefmark{3}%
            DEVCOM Army Research Lab, USA.
            lance.m.kaplan.civ@army.mil
    }
    \IEEEauthorblockA{
        \IEEEauthorrefmark{4}%
            ECE Department,
            University of California, Los Angeles, USA.
            mbs@ucla.edu
    }
    \IEEEauthorblockA{
        \IEEEauthorrefmark{6}%
            Air Force Research Laboratory, USA. 
            \{trevor.bihl.2, erik.blasch.1, nandini.iyer.2\}@us.af.mil
    }
}

\maketitle
\thispagestyle{fancy}

\begin{abstract}
\wifi devices, akin to passive radars, can discern human activities within indoor settings due to the human body's interaction with electromagnetic signals. Current \wifi sensing applications predominantly employ data-driven learning techniques to associate the fluctuations in the physical properties of the communication channel with the human activity causing them. However, these techniques often lack the desired flexibility and transparency.
This paper introduces \deepprobhar, a neuro-symbolic architecture for \wifi sensing, providing initial evidence that \wifi signals can differentiate between simple movements, such as leg or arm movements, which are integral to human activities like running or walking. The neuro-symbolic approach affords gathering such evidence without needing additional specialised data collection or labelling.
The training of \deepprobhar is facilitated by declarative domain knowledge obtained from a camera feed and by fusing signals from various antennas of the \wifi receivers.
\deepprobhar achieves results comparable to the state-of-the-art in human activity recognition. Moreover, as a by-product of the learning process, \deepprobhar generates specialised classifiers for simple movements that match the accuracy of models trained on finely labelled datasets, which would be particularly costly.
\end{abstract}


\section{Introduction}
\label{sec:introduction}
\wifi devices can be used as {\it passive radars} to recognise specific human activities in indoor environments because of the physical interaction of the human body with communication signals~\cite{li2020passive}. 
In \wifi, the \gls{csi} is a complex-valued vector computed at the receiver for every incoming frame that measures the wireless channel's properties and equalises the received signal.
However, the \gls{csi} also provides an electromagnetic fingerprint of the environment.

\begin{figure}
    \centering
    \includegraphics[width=\columnwidth]{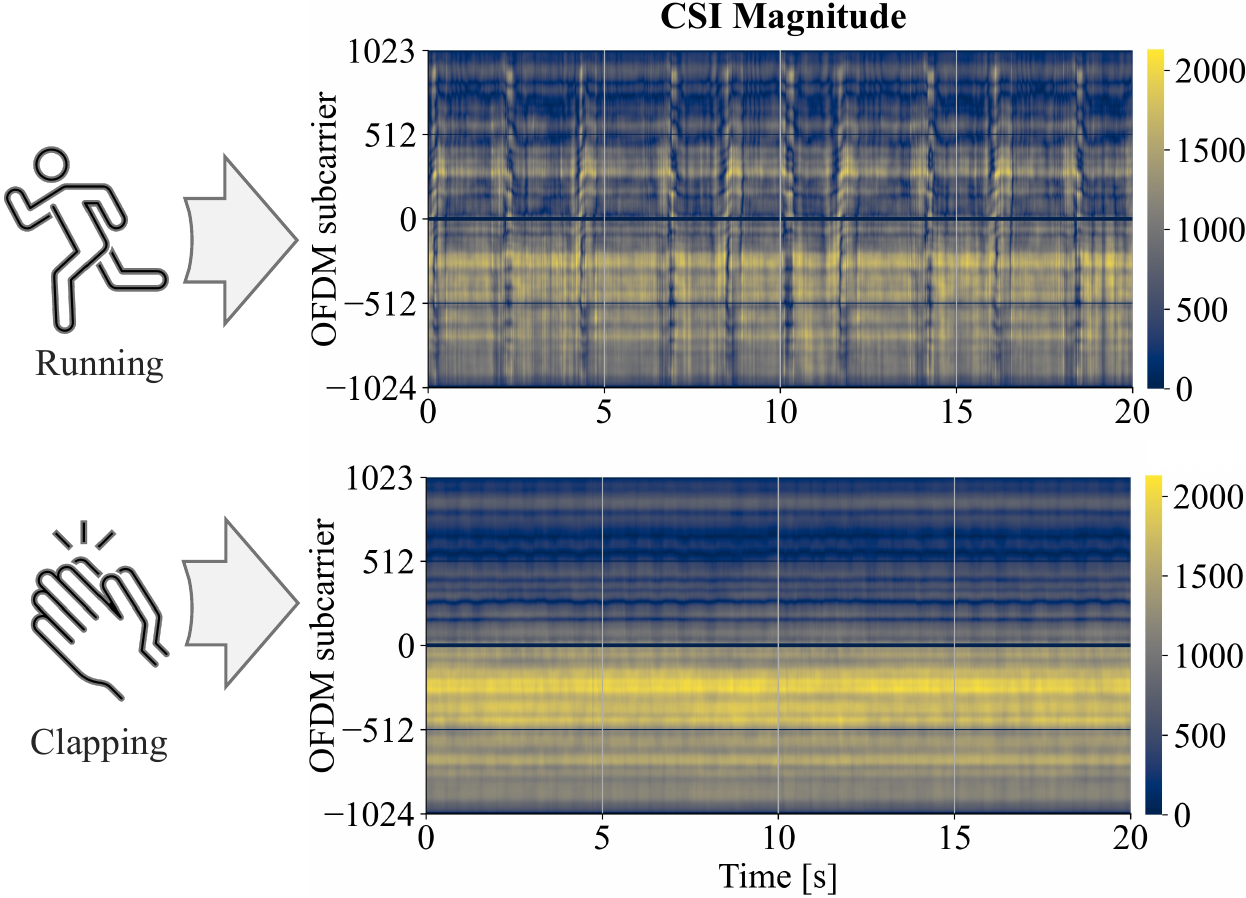}
    \caption{Magnitude of the \acrshort{csi} collected by the same antenna when a person performs two different activities, namely running (top) and clapping (bottom). \acrshort{csi} values are dimensionless and are reported as measured by the \wifi chipset.}
    \label{fig:csi_sample}
\end{figure}

\Cref{fig:csi_sample} (top) illustrates a snippet of the \gls{csi} captured by one single antenna while a person runs.
As the person moves around the room, the environment's effect on the signal changes due to the varying scattering on the human body.
The result is captured in a \emph{spectrogram} that highlights how the relative intensity of the signal changes over time and frequency.
The fundamental assumption of \gls{csi}-based \gls{har} is that it is possible to trace these variations back to the human activity that caused them, and in particular, to distinguish different types of activities, like running instead of standing still and clapping, \cf \cref{fig:csi_sample} (bottom).
However, such \wifi sensing applications employ techniques that often lack the desired flexibility and transparency.

In this paper, we introduce \deepprobhar, a neuro-symbolic approach to \gls{har} using a passive \wifi radar, providing initial evidence that \wifi signals can differentiate between simple movements, such as leg or arm movements, which are integral to human activities like running or walking.
\deepprobhar builds on top of a recent \cite{fusion2023} pre-processed dataset of human activities sensed through commercial \wifi devices \cite{exposingthecsi2023}, that makes use of \glspl{vae} \cite{kingma_AutoEncodingVariationalBayes_14} to identify a generative latent distribution seen as a compressed view of the original \gls{csi} signal.
Specifically, the paper discusses (\Cref{sec:background}) the differences between the two main approaches to \gls{har}: declarative and data-driven. Declarative approaches provide classification rules for defining activities but struggle with unstructured data \--- indeed, to our knowledge, they have not been proposed for \gls{csi} data; while data-driven approaches handle complex data types but are less flexible and more opaque. The paper expands on relevant references to the literature concerning data-driven approaches, including a description of our previous work published in \cite{fusion2023}, which provides a principled way to compress the \gls{csi} data and several architectures to fuse the signals coming from the different \wifi antennas of the passive radar.

A third method for \gls{har}, provided by neuro-symbolic approaches (\Cref{sec:methodology}), combines symbolic (declarative) reasoning techniques and \gls{nn} methods, aiming to improve the performance of AI systems~\cite{sheth2023neurosymbolic}.
Neuro-symbolic systems can merge the approximation capabilities of \glspl{nn} with the abstract reasoning abilities of symbolic methods, enabling them to extrapolate from limited data and produce interpretable results.
In particular, we extract declarative knowledge (a decision tree) of the human activities from a video feed captured by a camera observing the same environment as the \wifi receiver senses.
We then transfer such knowledge to train with just the label of the activity \deepprobhar, a neuro-symbolic architecture seeing different data modality (the \gls{csi}) and that builds on top of DeepProbLog \cite{manhaeve2018deepproblog,manhaeve2021neural}.

Our experimental results (\Cref{sec:results}) demonstrate that \deepprobhar achieves comparable if not better results than the state-of-the-art approaches while at the same time reaching a higher degree of transparency.
In particular, \deepprobhar can identify occurrences of simple movements (such as \textit{moving the upper arm}) without requiring specific labelling for such concepts.
Finally, \Cref{sec:discussion} highlights the ability to distinguish simple movements of the subject.

\section{Background}
\label{sec:background}
\subsection{Activity Recognition Approaches}
\label{sec:harapproaches}

There are two primary approaches to activity recognition: \textit{declarative} and \textit{data-driven}.
Declarative approaches~\cite{storf2009rule,theekakul2011rule,atzmueller2018explicative} provide classification rules that can be utilised to define the activity. An example of such a rule \--- in natural language \--- could be: \textit{running is the rapid, alternating action of pushing off and landing on the ground with one's feet}.\footnote{Microsoft Copilot on 14th March 2024.} However, the types of input data declarative approaches can handle are often limited.
Specifically, declarative rules typically require direct processing of the input data, which can pose challenges for unstructured data such as \wifi \gls{csi} (further discussed in \Cref{ssec:dataset}).
Indeed, the authors are unaware of any declarative approaches for \gls{har} operating over \gls{csi} data.

Data-driven approaches (\eg \cite{meneghello2022,bahadori2022rewis,liu2020,fusion2023}) are specifically designed to handle data types for which it is difficult to define rules directly. Despite their advantages, these approaches are more opaque and less flexible than their declarative counterparts, often making it impossible for the system's end-user to define patterns entirely.
Indeed, several \gls{har} systems work by deriving some physically-related quantity from some sensors (\eg the \glspl{csi}) that is then used to train a deep learning classification system \cite{meneghello2022,bahadori2022rewis,liu2020}.
In a previous work, we showed a principled approach to \gls{har} using a \gls{vae} generative model to compress the sensors' information and various architectures for fusing multiple antennas' signals~\cite{fusion2023}.

\subsection{Dataset}
\label{ssec:dataset}

In this work, we rely on a \gls{csi} dataset publicly released by members of the author list.\footnote{https://github.com/ansresearch/exposing-the-csi}
Further details about the dataset are available in~\cite{exposingthecsi2023}.
The experimental testbed comprises two Asus RT-AX86U devices placed on opposite sides of an approximately 46-square-metre room.
One device generates dummy IEEE~802.11ax (\wifi~6) traffic at a constant rate of 150 frames per second using the frame injection feature in \cite{axcsi2021}.
The other device (also called \emph{monitor}) receives the \wifi frames and stores the associated \gls{csi} for each of its four receiving antennas independently.
Meanwhile, one candidate performs different activities in the middle of the room.

\begin{figure}
    \centering
    \includegraphics[width=0.95\columnwidth]{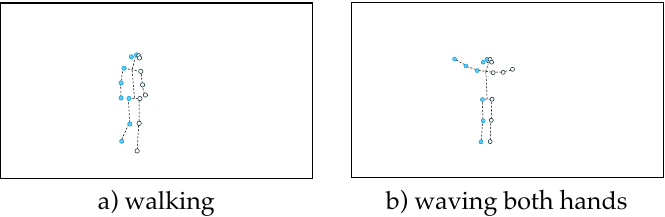}
    \caption{Sample of the video dataset for two different activities: a) {\it walking} and b) {\it waving both hands}. The key points in every video frame help to discern the right side (highlighted with coloured dots) from the left side of the candidate.}
    \label{fig:videopose3d}
\end{figure}

The \gls{csi} dataset is coarsely synchronised with a video recording of the activities, collected using a smartphone camera placed in a fixed location and then anonymised.
Specifically, to preserve the participants' identity, VideoPose3D~\cite{videopose3d} was used to extract a model of the candidate performing the activities.
VideoPose3D identifies 17 key points to track the motion of the main human joints, as shown in \cref{fig:videopose3d}.
The key points are stored as a list of $(x,y)$ coordinates in the camera viewport for each video frame.
Even though the dataset includes the \gls{csi} data of twelve different activities, there are seven activities in total for which both \gls{csi} and video data are available: \emph{walking, running, jumping, squatting, waving both hands, clapping,} and \emph{wiping}.
For each activity, the dataset contains 80 seconds of \gls{csi} data (sampled at 150 \gls{csi} per second) and the corresponding video data (\ie the key points of the candidate, sampled at 30~fps).
VideoPose3D can also reconstruct a 3D model of the candidate using a deep learning algorithm; however, we found some numerical instability in the 3D coordinates reconstructed by the tool.
Hence, in this work, we only consider the 2D coordinates of the joints extracted from the original video traces.

\subsection{Dataset Pre-Processing and VAE Architectures}
\label{sec:vae-architectures}

\begin{figure}
    \centering
    \includegraphics[width=0.82\columnwidth]{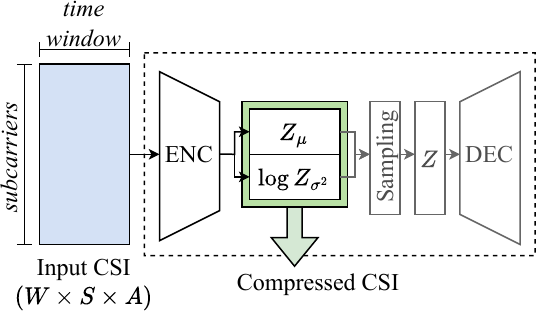}
    \caption{The \acrshort{vae} models map the input \acrshort{csi} data onto the four parameters of a bivariate Gaussian distribution (mean $Z_\mu$ and variance $Z_{\sigma^2}$ along two axes), which we can be used as a compressed representation of the input \acrshort{csi}.}
    \label{fig:vae-architecture}
\end{figure}

The work in \cite{fusion2023} introduced several modular architectures for \gls{har} using \gls{csi} data, practically splitting the problem into two separate sub-tasks.
First, a \gls{vae} provided a concise (yet informative) characterisation of the different activities as perceived by the \wifi monitor through the \gls{csi}.
Specifically, the \gls{vae} mapped short sequences of \gls{csi} data \--- sampled using a sliding window in the time domain \--- onto a latent bivariate Gaussian distribution defined by 4 parameters (\ie mean and variance along two dimensions).
Second, a \gls{mlp} trained on the latent space parameters of the \gls{vae} was used to classify the different activities.
The unsupervised training of the \gls{vae} can be carried out separately from the training of the \gls{mlp}.
However, since the \wifi monitor is sensing the environment using four physically-spaced antennas, several \gls{vae} architectures were proposed in \cite{fusion2023} to evaluate different strategies to fuse the information sensed by every antenna.
\Cref{fig:vae-architecture} summarises the general architecture of the \glspl{vae}.
The input data structure is a tensor of size $(W \times S \times A)$, where $W$ is the number of \gls{csi} samples in a given time window, $S$ is the number of subcarriers in a \wifi frame, and $A$ is the number of antennas considered by the \gls{vae}.
We fixed the time window to 3~seconds, so $W$=450, while the considered dataset contains \wifi~6 frames with $S$=2048 subcarriers (\mbox{160-MHz} bandwidth).

In this work, we start from the compressed representation of the \gls{csi} data windows in the \gls{vae}'s latent space to develop a neuro-symbolic architecture for \gls{har}.
This pre-processed dataset was obtained by training the \glspl{vae} described in \cite{fusion2023} on the complete set of activities in the original dataset.
Here, we briefly report the resultant architectures for convenience.

First, we consider a set of architectures called \boldnofusing{x}.
These architectures have been trained using the data incoming from one single antenna of the \wifi monitor ($A$=1 in \cref{fig:vae-architecture}).
We denote with the letter x the antenna of the monitor whose data we used to train the \gls{vae}.
Hence, we define four separate architectures, one for each antenna: \boldnofusing{1}, \boldnofusing{2}, \boldnofusing{3}, and \boldnofusing{4}.

While using data from one single antenna can be enough for some \gls{har} applications~\cite{rfnet2020}, we proved in previous work that there are consistent advantages in fusing the \gls{csi} data from different antennas~\cite{fusion2023}.
Therefore, we also consider a second type of architecture, called \boldearlyfusing.
In this case, the \gls{csi} data from all four monitor antennas are stacked together in the input data structure ($A$=4 in \cref{fig:vae-architecture}).
Still, the latent space of \vaef has a bivariate latent normal distribution which can condense together cross-antennas regularities.

The last architecture we consider, called \bolddelayedfusing, employs all the four \glspl{vae} trained independently on each monitor antenna and concatenates their latent space parameters into a single vector of 16 elements (4 features for each \gls{vae}) that becomes the input of the following classification stage.

\section{Methodology}
\label{sec:methodology}
In this section we present \deepprobhar, the first neuro-symbolic approach to human activity recognition fusing information from \wifi \glspl{csi}.
First, we briefly introduce neuro-symbolic AI (\cref{sec:deepproblogintro}), particularly the DeepProbLog approach~\cite{manhaeve2018deepproblog,manhaeve2021neural}.
Then, we discuss how we extracted domain-dependent knowledge for classifying different activities using a more interpretable modality, \viz the video recording of the performed activities (\cref{sec:ruleextraction}).
For this work, we relied on such a knowledge extraction to reasonably assume that the \wifi sensor could have captured the way we would describe activities, as both the camera and the antennas were looking at the same environment.
Finally, we describe in detail the \deepprobhar architecture (\cref{sec:proposal}).

\subsection{Primer on Neuro-Symbolic AI: the DeepProbLog Approach}
\label{sec:deepproblogintro}

Neurosymbolic AI, \eg \cite{sheth2023neurosymbolic}, is often referred to as the combination of symbolic reasoning techniques and neural network methods to improve the performance of AI systems. These systems can merge the robust approximation capabilities of neural networks with the abstract reasoning abilities of symbolic methods, enabling them to extrapolate from limited data and produce interpretable results.
Neurosymbolic AI techniques can be broadly categorised into two groups. The first considers techniques that condense structured symbolic knowledge for integration with neural patterns and reason using these integrated neural patterns. The second considers techniques that extract information from neural patterns to facilitate mapping to structured symbolic knowledge (\ie \textit{lifting}) and carry out symbolic reasoning. 

This paper focuses on a specific approach within the second group of neurosymbolic AI techniques, \viz DeepProbLog \cite{manhaeve2018deepproblog,manhaeve2021neural}. To present it, we first need to briefly introduce ProbLog, \cite{de2007problog}, which is a probabilistic logic programming language. A ProbLog program comprises a collection of probabilistic facts, denoted as $F$, and a set of rules, denoted as $R$. Facts are expressed in the form $p::f$, where $f$ is an atom symbolising a notion that can either be true or false. $p$ is a probability value ranging from 0 to 1, which signifies the probability of the fact being true. Rules are expressed in $h \leftarrow b_1,...,b_n$, where $h$ is an atom and $b_i$ are literals. A literal can be an atom or the negation of an atom.

ProbLog includes \glspl{ad} as a syntactic extension of the form
$p_1 :: h_1; ... ; p_n :: h_n \leftarrow b_1, ..., b_m.$ 
where the $p_i$ are probabilities such that $\sum p_i = 1$, and $h_i$ and $b_j$ are atoms. 
It is immediate to see that they encode categorical distributions over the possible results of a random variable that can take on one of $K$ possible categories. Moreover, it is also a way to capture a Bernoulli distribution where we wish to name the two outcomes explicitly, \eg \textit{head} and \textit{tail} as possible results of tossing a coin, rather than limiting ourselves to just one, let us say \textit{head}, and deriving the other \--- \ie \textit{tail} \--- as the negation.

A Problog program can be encoded in a probabilistic circuit \cite{fierens2015inference}, which is a graphical model that compactly represents probability distributions. Each fact and rule in a Problog program can be translated into a node or a set of nodes in a probabilistic circuit, and the probabilities associated with the facts correspond to the parameters of the probabilistic circuit.

DeepProbLog \cite{manhaeve2018deepproblog,manhaeve2021neural} is a programming language that combines neural networks with probabilistic logic. It extends ProbLog by introducing \glspl{nad}. Differently from \glspl{ad}, in \glspl{nad} the probabilities of the categorical distribution are the output layer of a neural network $f(\bm{x}, \bm{\theta})$. There are no restrictions on the form of $f(\cdot)$ as long as it outputs a categorical distribution over $K$ classes, \eg using a \textit{softmax} activation function at the network output. Each \gls{nad} is thus associated to a specific neural network $f(\cdot)$

For each \gls{nad}, DeepPropLog computes the gradient of the loss w.r.t. the output of the associated $f(\bm{x}, \bm{\theta})$. Standard backpropagation algorithms use the gradient to train the parameters $\bm{\theta}$. Such a computation leverages the differentiability of the ProbLog program, the computational machinery of which can be expressed over the associated probabilistic circuits. For further details, the interested reader is referred to \cite{manhaeve2021neural} and for applications of DeepProbLog to analogous tasks such as complex event processing, to \cite{vilamala2023deepprobcep}.

\begin{figure}
    \centering
    \includegraphics[width=0.66\columnwidth]{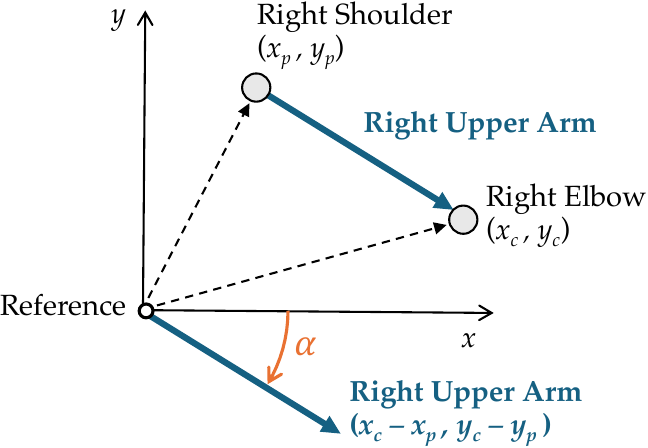}
    \caption{Computation of the right upper arm's angle \limbangle. The same operation applies to all the other limb segments.}
    \label{fig:trigonometry}
\end{figure}

\begin{figure}
    \centering
    \includegraphics[width=0.94\columnwidth]{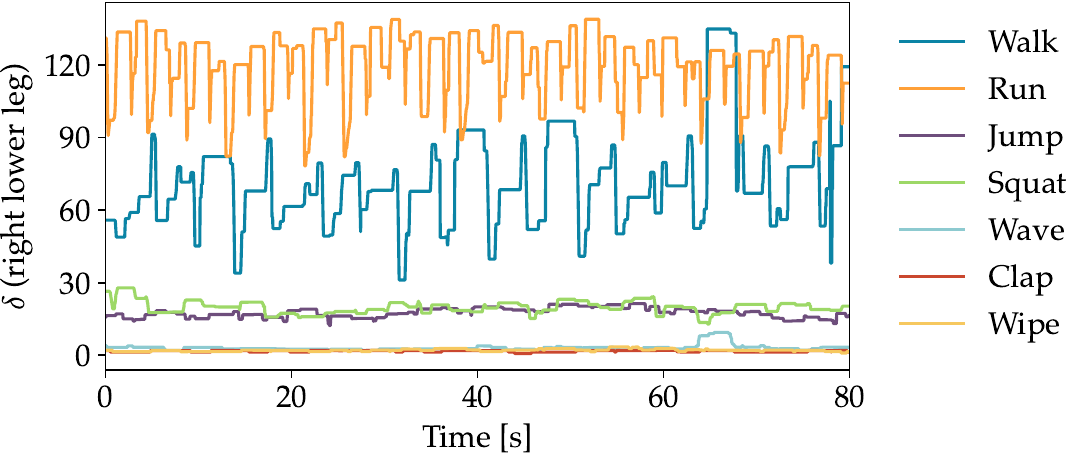}
    \caption{The feature $\anglefeature_l$ corresponding the right lower leg indicates the motion of that limb for each target activity.}
    \label{fig:feature-example}
\end{figure}

\begin{figure}
    \centering
    \includegraphics[width=\columnwidth]{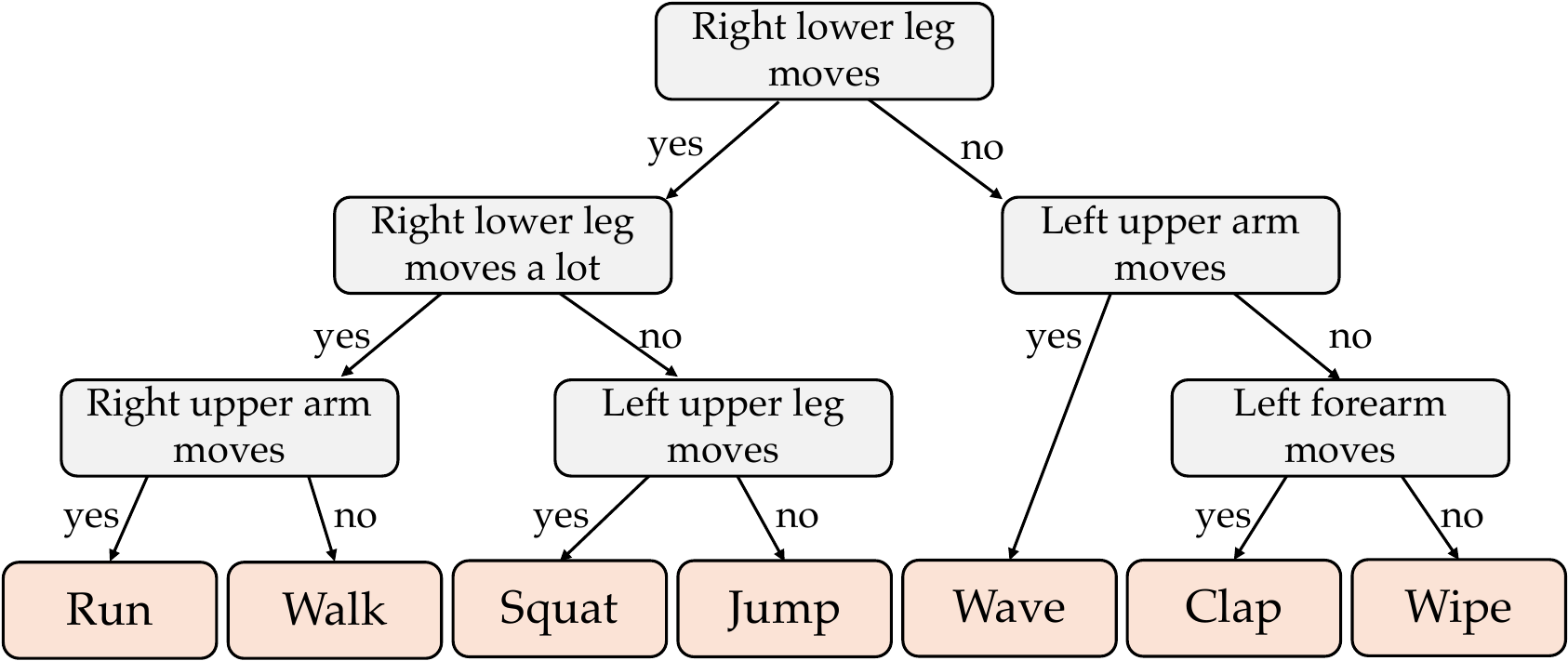}
    \caption{Decision tree derived from the video data analysis. Leaf nodes represent the target activities. At every decision node, a single feature $\anglefeature_l$ is tested against a threshold to determine whether the corresponding limb is moving.}
    \label{fig:decision_tree_video}
\end{figure}

\subsection{Domain-Dependent Rule from Different Modality}
\label{sec:ruleextraction}

To operate using a neuro-symbolic approach, we must use some declarative knowledge to describe the activities we plan to classify. 
We assume that every target activity can be defined by combining basic movements; these ``atomic'' movements should be sufficiently easy to identify.
For instance, \emph{running} mandates a rapid alternate motion of the legs accompanied by broad arms' movement; conversely, \emph{clapping} can be defined by the repetitive motion of the forearms.
Building on this assumption, we can combine a limited subset of basic movements into a potentially large set of target activities.

In the following, we describe the process of extracting indicators of simple movements from the video dataset introduced in \cref{ssec:dataset}.
The idea is to compute a vector of scalar values (one for each limb) that characterise the \emph{amount of motion} of the limbs in a given time window.
Such values will serve as numerical features that can be combined and used to describe more complex target human activities.

A brief analysis of the video dataset revealed that we have access to the $(x,y)$ coordinates of 17 key points in the camera viewport for each video frame.
Specifically, 12 key points correspond to the person's shoulders, elbows, hands, hips, knees, and feet (one for each side, \cf \cref{fig:videopose3d}).
Given the location of the joints and using basic trigonometry, we can precisely locate the limbs' position and orientation in the 2D frame.
We consider $L=8$ limb segments: two upper arms, lower arms, upper legs, and lower legs.
Each segment is defined by two joints, a parent joint $p$ (\eg the shoulder for the upper arm limb) and a child joint $c$ (\eg the elbow for the upper arm).
\Cref{fig:trigonometry} shows that if a parent joint $p$ has coordinates $(x_p, y_p)$ and a child joint $c$ has coordinates $(x_c, y_c)$, then the limb segment vector draws an angle
$\limbangle = \arctan \left( \frac{y_c-y_p}{x_c-x_p} \right)$
which changes at every frame depending on its motion.

\begin{figure}
    \centering
    \includegraphics[width=0.76\columnwidth]{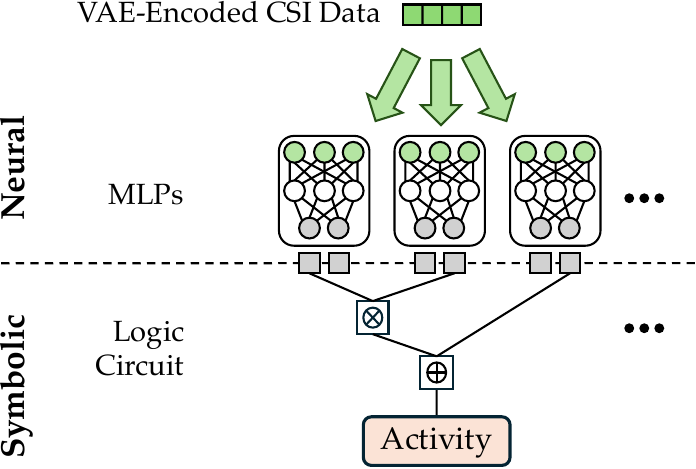}
    \caption{Architecture of \deepprobhar. The neural part extracts \emph{concepts} from the compressed \gls{csi}, while a logic circuit combines such \emph{concepts} to derive the target activity.}
    \label{fig:ns-architecture}
\end{figure}

\begin{figure*}[t]
    \centering
    \begin{subfigure}[b]{0.6\linewidth}
         \centering
         \includegraphics[width=\textwidth]{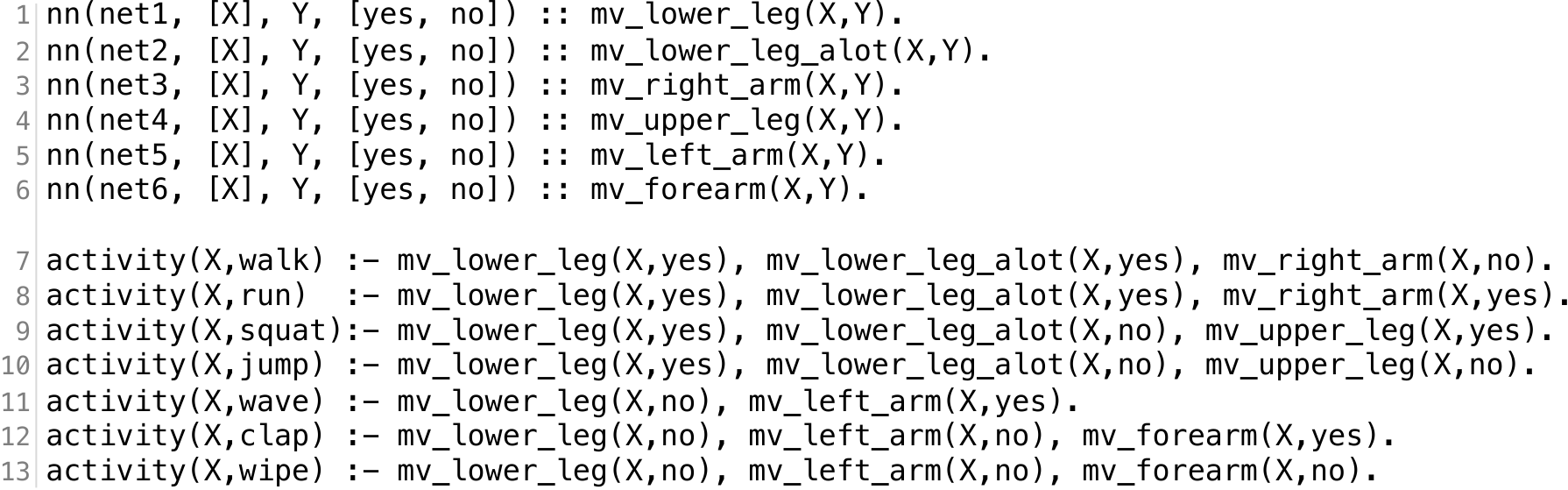}
         \caption{The DeepProbLog program.}
         \label{fig:deepproblog-program}
    \end{subfigure}
    \hfill
    \begin{subfigure}[b]{0.39\linewidth}
         \centering
         \includegraphics[width=0.7\textwidth]{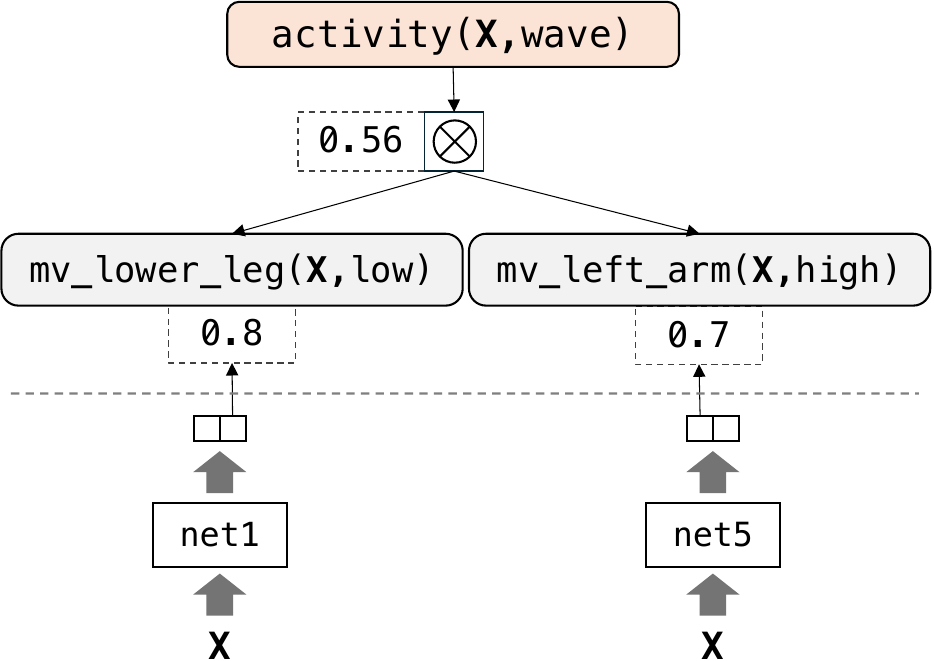}
         \caption{Logic circuit for the query \texttt{activity(X,wave)}.}
         \label{fig:sample-query}
    \end{subfigure}
    \caption{(a) DeepProbLog code implementing the neuro-symbolic architecture and (b) representation of the logic circuit for the query \texttt{activity(X,wave)}. Notice we drop the left/right distinction when features are uniquely defined in the code.}
\end{figure*}

We index the different limb angles with $\limbangle_l$, where $l$ can range from 1 to $L$.
To quantify the motion of each limb, we consider a sliding time window of duration $T=3$ seconds, matching the time window already applied on the \gls{csi} data processed by the \glspl{vae}.
Since the video was recorded at 30~fps, every time window contains 90 samples of the angles $\limbangle_l$.
Then, for each time window $t$, we compute the dynamic range of every angle $\limbangle_l$.
The result is a feature vector $\boldsymbol{\anglefeature}(t)$ with $L$ elements describing the amount of motion of each limb during a time window.
Let us assume that a limb was not moving during the time window $t$ (\eg one leg when the candidate is clapping); then, the associated $\limbangle_l$ is barely changing, and $\anglefeature_l(t)$ will be minimal.
On the other hand, if we look at the same leg when the person is walking, we will see that $\limbangle_l$ describes broader trajectories, and $\anglefeature_l(t)$ will be much more significant.

The computation of the dynamic range of the angles $\limbangle_l$ provides a quick estimation of each limb's motion amount. It is a simple way to discriminate the target activities in our dataset.
In \cref{fig:feature-example}, we show how just the feature $\anglefeature_l$ (where $l$ is the right lower leg) is enough to separate three families of activities.
The first family involves broad movements of the lower legs ({\it walking} and {\it running}); the second one involves modest movements of the lower legs ({\it jumping} and {\it squatting}); finally, the last family of activities consists of no movement of the leg at all ({\it waving}, {\it clapping}, and {\it wiping}).
Arguably, the limbs' motion can be estimated using more sophisticated approaches, \eg involving a Fourier analysis of the angles $\limbangle_l$ to identify periodic patterns or taking into account the perspective effects.
However, we argue that taking the dynamic range is a reasonably simple and effective solution.

The video dataset analysis and the features extracted from the limbs' motion resulted in the decision tree shown in \cref{fig:decision_tree_video}.
This rule-based classification combines six \emph{concepts} extracted from the dataset that can take binary values.
At every decision node, one single value of the $\boldsymbol{\anglefeature}$ vector is compared against a threshold that determines whether the corresponding limb is moving.
Note that, in general, we are not bound to consider just binary decisions; indeed, according to what we already observed in \cref{fig:feature-example}, the first node at the top of the tree in \cref{fig:decision_tree_video} could have had three outputs.
However, we forced the model to have only binary decisions to simplify the implementation of our neuro-symbolic architecture.
In this way, the same feature is tested twice against two different thresholds (\cf, {\it Right lower leg moves} and {\it Right lower leg moves a lot} in \cref{fig:decision_tree_video}).

\subsection{\deepprobhar: A Neuro-Symbolic Architecture for Human Activity Recognition Using Wi-Fi Data}
\label{sec:proposal}

In \cref{sec:ruleextraction}, we have defined the declarative knowledge necessary to identify the different target activities and a simple way to extract the information from the video data.
This section introduces \deepprobhar, the first neuro-symbolic system designed explicitly for \gls{har} applications using \wifi sensing data.

A schematic overview of the proposed architecture is shown in \cref{fig:ns-architecture}.
The system takes in input the compressed \gls{csi} data encoded by one of the \glspl{vae} introduced in \cref{sec:vae-architectures}.
This implies that we can define several architectures depending on the \gls{vae} used to pre-process the dataset.
We consider four single-antenna architectures \nofusing{x} ($\mathrm{x}$ ranging from 1 to 4) and two architectures fusing the data from multiple antennas, namely \earlyfusing and \delayedfusing (\cf \cref{sec:vae-architectures}).
All the \deepprobhar architectures employ six small \glspl{mlp} to extract binary symbols from the input \gls{csi} data, which are combined using logic rules to estimate the target activity.
Every \gls{mlp} contains only two hidden layers with 8 neurons each, activated using a ReLU function, and a binary output layer with a SoftMax activation function.

The DeepProbLog code used to combine the neural networks' output with the logic part of the architecture is listed in \cref{fig:deepproblog-program}.
The DeepProbLog program defines the six \glspl{mlp} such that each \gls{mlp} should correspond to a different decision node in \cref{fig:decision_tree_video}.
Then, a list of predicates implements the rules described in the decision tree constructed starting from the video data analysis.
\Cref{fig:sample-query} shows an example of the logic circuit derived from the DeepProbLog code, where the grey rectangles correspond to the probabilistic facts identified by the neural networks \texttt{net1} and \texttt{net5} and the red rectangle corresponds to the query defined by the formula on line 11 of the code listing.
The white box with the $\otimes$ symbol represents the logical operator \texttt{AND} applied to its children.

\section{Experimental Results}
\label{sec:results}
We now evaluate the classification performance of all the \deepprobhar models on the selected public dataset, which comprises both the \gls{csi} data and anonymised video recordings~\cite{exposingthecsi2023}.
First, we certify that the rules extracted from the video dataset yield good accuracy in estimating the target activities (\cref{sec:validation-video}).
Then, we measure the classification accuracy of the \deepprobhar models (\cref{sec:performance}).
Since \deepprobhar leverages declarative knowledge gathered for a different modality (the video data feed), we expect such a reference to be the upper limit that \deepprobhar can reach.
Finally, we compare the results obtained with the neuro-symbolic architecture with those obtained by more traditional approaches based on neural networks\footnote{Code available: \url{https://github.com/marcocominelli/csi-vae/tree/fusion2024}}.

\begin{table}[t]
    \centering
    \caption{Accuracy and average precision, recall and F1 score of \deepprobhar with different data fusion strategies. The accuracy of the extracted declarative knowledge tested over the video feed is provided as \textit{Video reference} and is the upper limit of \deepprobhar's performance.}
    \begin{tabular}{lcccc}
        \toprule
        {\bf Architecture} & {\bf Accuracy} & {\bf Precision} & {\bf Recall} & {\bf F1} \\
        \midrule
        {\it Video reference} & {\it 0.98} & {\it 0.99} & {\it 0.98} & {\it 0.98} \\
        \midrule
        No-Fused-1 & 0.50 & 0.49 & 0.51 & 0.50 \\
        No-Fused-2 & 0.62 & 0.62 & 0.63 & 0.62 \\
        No-Fused-3 & 0.53 & 0.53 & 0.51 & 0.52 \\
        No-Fused-4 & 0.76 & 0.76 & 0.77 & 0.77 \\
        Early-Fusing & 0.84 & 0.84 & 0.85 & 0.84 \\
        Delayed-Fusing & {\bf 0.95} & {\bf 0.95} & {\bf 0.95} & {\bf 0.95} \\
        \bottomrule
    \end{tabular}
    \label{tab:main_results}
\end{table}

\begin{table}[t]
    \centering
    \caption{Comparison with state-of-the-art non-neuro-symbolic approaches using a single MLP trained on the corresponding dataset of each different architecture.}
    \begin{tabular}{lccc}
        \toprule
        {\bf Architecture} & {\bf \deepprobhar} & {\bf Small MLP} & {\bf Large MLP} \\
        \midrule
        No-Fused-1 & 0.50 & 0.59 & 0.62 \\
        No-Fused-2 & 0.62 & 0.74 & 0.76 \\
        No-Fused-3 & 0.53 & 0.58 & 0.59 \\
        No-Fused-4 & 0.76 & 0.78 & 0.80 \\
        Early-Fusing & 0.84 & 0.88 & 0.89 \\
        Delayed-Fusing & {\bf 0.95} & {\bf 0.94} & {\bf 0.98} \\
        \bottomrule
    \end{tabular}
    \label{tab:mlp_comparison}
\end{table}

\subsection{Validation of Declarative Knowledge}
\label{sec:validation-video}

To ensure that the declarative knowledge gathered from the video data is enough to produce sensible guesses about the target activity, we implemented the rule-based classifier introduced in \cref{fig:decision_tree_video}, using manually fine-tuned thresholds on the angles' features ${\boldsymbol{\anglefeature}}$) extracted directly from the video data.
Such a classifier operating over the video feed achieves \emph{an accuracy of 98\%} over the seven target activities.
Interestingly, classification errors happen between the activities \emph{walk} and \emph{run}.
Arguably, these are the most challenging activities to discriminate, even for a human viewer, mainly because of the indoor experimental setting.

\begin{table*}
    \centering
    \caption{Classification accuracy of the various \deepprobhar's MLPs compared to specialised MLPs trained on a finely labelled dataset of simple movements.}
    \begin{tabular}{lccccccc}
        \toprule
        \multicolumn{8}{l}{\bf \deepprobhar's MLPs} \\
         & {\bf MLP 1} & {\bf MLP 2} & {\bf MLP 3} & {\bf MLP 4} & {\bf MLP 5} & {\bf MLP 6} & {\bf Overall} \\
         {\bf Architecture} & {\bf (right lower leg)} & {\bf (right lower leg \#2)} & {\bf (right arm)} & {\bf (left upper leg)} & {\bf (left arm)} & {\bf (left forearm)} & {\bf Accuracy} \\
        \midrule
        No-Fused-1 & 0.80 & 0.79 & 0.65 & 0.84 & 0.67 & 0.84 & 0.50 \\
        No-Fused-2 & 0.82 & 0.74 & 0.97 & 0.98 & 0.89 & 0.93 & 0.62 \\
        No-Fused-3 & 0.86 & 0.66 & 0.72 & 0.70 & 0.88 & 0.97 & 0.53 \\
        No-Fused-4 & 0.87 & \textbf{0.98} &  0.92 & 0.87 & 0.88 & {\bf 1.00} & 0.76 \\
        Early-Fusing & 0.94 & \textbf{0.98} & 0.85 & 0.99 & 0.88 & 0.99 & 0.84 \\
        Delayed-Fusing & {\bf 0.99} & {\bf 0.98} & {\bf 0.96} & {\bf 1.00} & {\bf 0.96} & {\bf 1.00} & {\bf 0.95} \\
        \bottomrule
        \toprule
        \multicolumn{8}{l}{\bf MLPs trained independently on finely labelled dataset of simple movements} \\
         & {\bf MLP 1} & {\bf MLP 2} & {\bf MLP 3} & {\bf MLP 4} & {\bf MLP 5} & {\bf MLP 6} & {\bf Overall} \\
         {\bf Architecture} & {\bf (right lower leg)} & {\bf (right lower leg \#2)} & {\bf (right arm)} & {\bf (left upper leg)} & {\bf (left arm)} & {\bf (left forearm)} & {\bf Accuracy} \\
        \midrule
        No-Fused-1 & 0.83 & 0.81 & 0.68 & 0.84 & 0.85 & 0.90 & 0.59 \\
        No-Fused-2 & 0.87 & 0.82 & 0.97 & 0.99 & 0.90 & 0.93 & 0.70 \\
        No-Fused-3 & 0.89 & 0.68 & 0.74 & 0.70 & 0.90 & 0.99 & 0.58 \\
        No-Fused-4 & 0.90 & 0.98 & 0.92 & 0.88 & 0.88 & {\bf 1.00} & 0.79 \\
        Early-Fusing & 0.96 & 0.98 & 0.87 & {\bf 1.00} & 0.89 & {\bf 1.00} & 0.86 \\
        Delayed-Fusing & {\bf 1.00} & {\bf 0.99} & {\bf 0.98} & {\bf 1.00} & {\bf 0.98} & {\bf 1.00} & {\bf 0.98} \\
        \bottomrule
    \end{tabular}
    \label{tab:independent_mlp_results}
\end{table*}

\subsection{Performance of \deepprobhar}
\label{sec:performance}

To evaluate the performance of the different architectures, we partition every compressed \gls{csi} dataset into a training and a testing set with an 80/20 split.
In the following, all the models are trained with a learning rate of 0.001 for 20 epochs.

\Cref{tab:main_results} summarises the main results of the \deepprobhar architectures for all the fusion strategies considered in \cite{fusion2023} (\cf \cref{sec:proposal}).
Similarly to \cite{fusion2023}'s results, the \delayedfusing fusion strategy yields the best results.
The model's accuracy that combines the data of different antennas, each processed by a separate \gls{vae}, closely approaches the accuracy of the reference classifier trained on video data.
However, as highlighted by the confusion matrixes of \deepprobhar for the various data fusion techniques (\Cref{fig:CM-main-results}), classification errors are not limited to the classes \emph{walk} and \emph{run}.

In \cref{tab:mlp_comparison}, we compare the results with two state-of-the-art non-neuro-symbolic architectures derived from the work in \cite{fusion2023}.
These architectures use the same \gls{vae} and one single \gls{mlp} substituting the entire neuro-symbolic architecture.
In the neuro-symbolic architecture, each of the six \glspl{mlp} learning a separate feature has 130 parameters (226 for the \delayedfusing approach).
If we try to approximate the neuro-symbolic architecture using a single \textit{small \gls{mlp}} (2 hidden layers, 8 neurons each), the resulting model has 175 parameters (271 for the \delayedfusing approach).
Arguably, since the neuro-symbolic architecture features six \glspl{mlp}, it would be interesting to consider a \textit{large \gls{mlp}} (2 hidden layers, 22 neurons each) whose number of parameters closely matches the one of all the neuro-symbolic \glspl{mlp}.
The results in \cref{tab:mlp_comparison} reveal that the neuro-symbolic architectures perform worse than the single \glspl{mlp} when considering just one antenna, but their accuracy becomes similar when fusing the data from multiple antennas.
We also highlight that the models in \cite{fusion2023} were evaluated on different activities, so the corresponding \glspl{mlp} have been trained from scratch in this work.

\section{Discussion}
\label{sec:discussion}
\Cref{tab:independent_mlp_results} (top) shows the results of the six \glspl{mlp} that have been trained in \deepprobhar, one for each feature as illustrated in \Cref{fig:decision_tree_video}. For the sake of comparison, we also trained independently six \gls{mlp} over a finely labelled dataset of simple movements \--- it is worth mentioning that labelling such a dataset could be extremely costly in less controlled settings \--- so that each \gls{mlp} was optimised to classify only one of the relevant features, see \Cref{tab:independent_mlp_results} (bottom).

We wish to point out three aspects.
First, the performance of each of the \deepprobhar's \gls{mlp} trained on sparse data (top of the table) appears to be close to the optimised \glspl{mlp} trained over the finely-labelled dataset.

Secondly, the overall accuracy (last column) is lower than each of the accuracies for all the \glspl{mlp}.
This is due to the independence assumptions underlying the training of all such neural networks and their subsequent usage for classification, whether via \deepprobhar (which relies on the same strong independence assumptions of ProbLog \cite{deraedt_Inducingprobabilisticrelational_15,fierens2015inference}\footnote{For a more comprehensive discussion on the role of probabilistic dependencies among variables in probabilistic circuits \--- including those derived from ProbLog \--- we refer the interested reader to \cite{cerutti2022handling}.}) or by a deterministic classifier that follows the decision tree in \cref{fig:decision_tree_video}. 
Indeed, given a neural network $f(\cdot)$, its accuracy can be seen as the probability of $f(\cdot)$ to return the correct answer for a given input.
In \cref{fig:decision_tree_video}, for instance, we see that classifying {\it running} and {\it walking} relies on three of the classifiers whose accuracies are available in \Cref{tab:independent_mlp_results}: {\bf MLP 1} that tells if the right lower leg moves; {\bf MLP 2} that tells if the right lower leg moves a lot; and {\bf MLP 3} that tells if the right upper arm moves.
Under independence assumptions, the probability of correct classification of {\it running} and {\it walking} is the product of the probability that each of the three classifiers returns a correct answer.
The average of the probabilities of correct classification for each of the activities as the product of the probabilities of correct classifications for the \glspl{mlp} used for such a classification according to \Cref{fig:decision_tree_video} amounts to the same overall accuracy as computed in the last column of \Cref{tab:independent_mlp_results}. 

\begin{figure}
    \centering
    
    \begin{subfigure}[b]{0.455\linewidth}
         \centering
         \includegraphics[width=\textwidth]{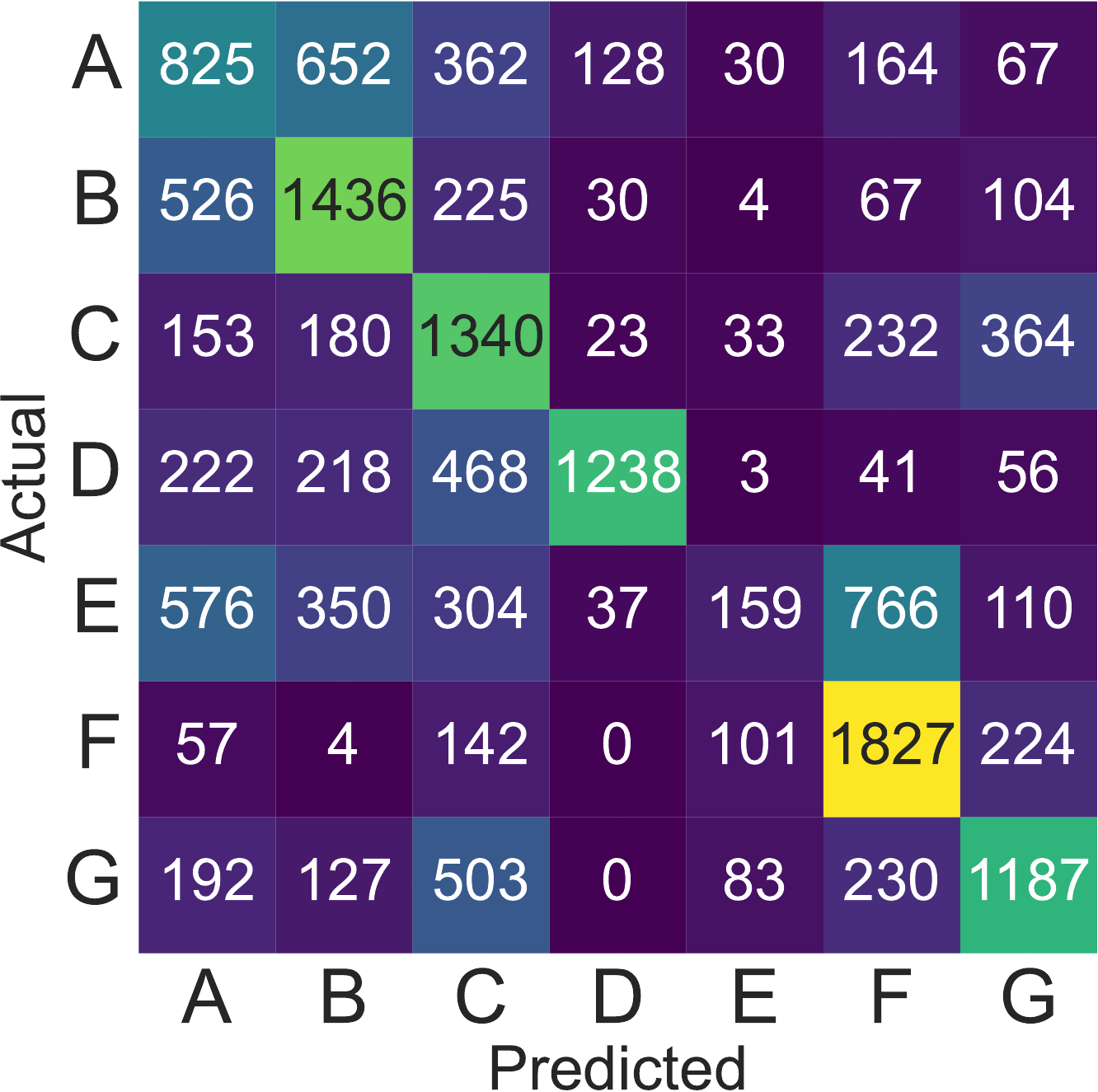}
         \caption{\nofusing{1}}
         \label{fig:cm-no-fused-1}
     \end{subfigure}
    \hfill
    \begin{subfigure}[b]{0.455\linewidth}
         \centering
         \includegraphics[width=\textwidth]{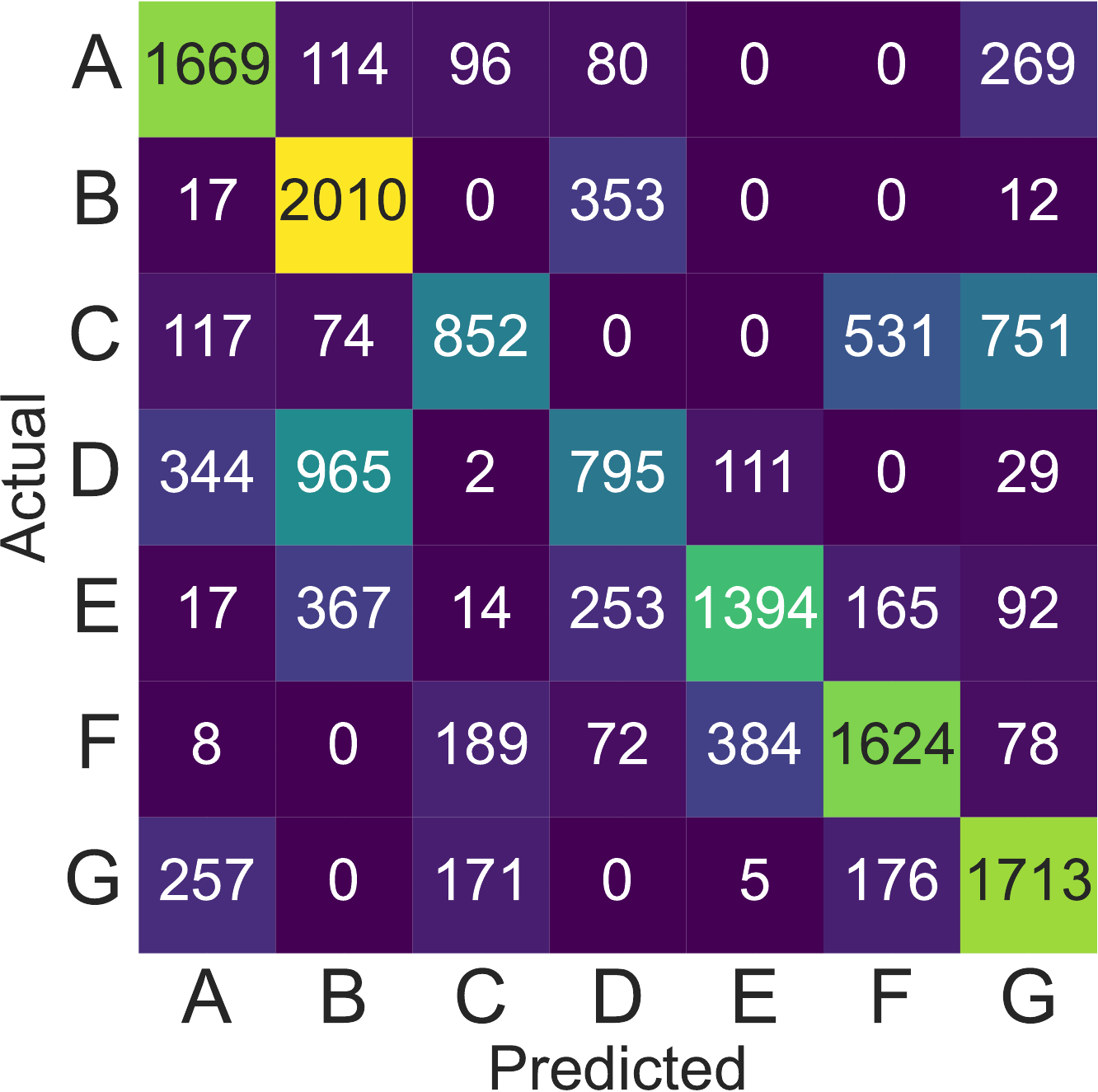}
         \caption{\nofusing{2}}
         \label{fig:cm-no-fused-2}
     \end{subfigure}
     
     \begin{subfigure}[b]{0.455\linewidth}
         \centering
         \includegraphics[width=\textwidth]{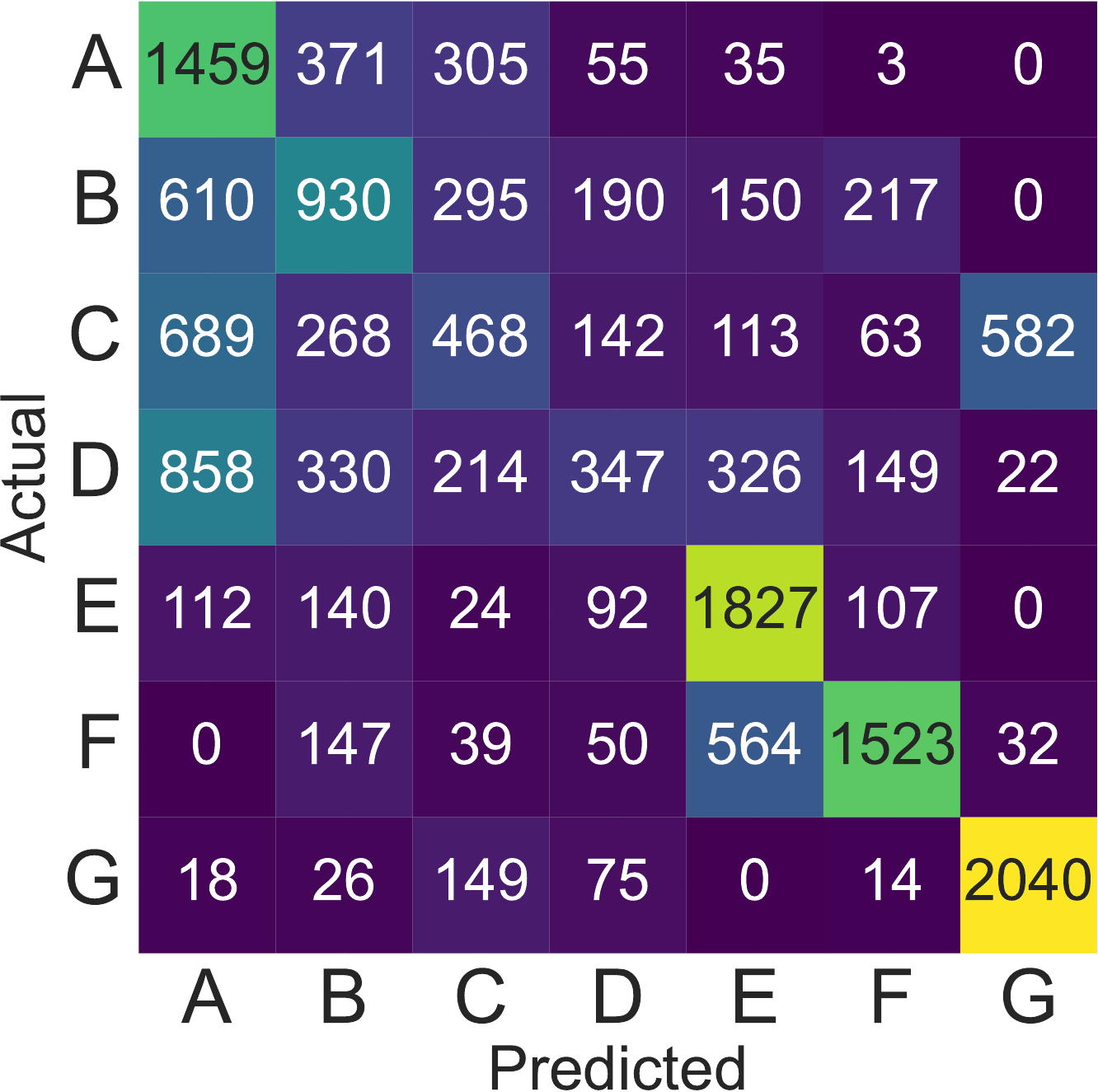}
         \caption{\nofusing{3}}
         \label{fig:cm-no-fused-3}
     \end{subfigure}
    \hfill
    \begin{subfigure}[b]{0.455\linewidth}
         \centering
         \includegraphics[width=\textwidth]{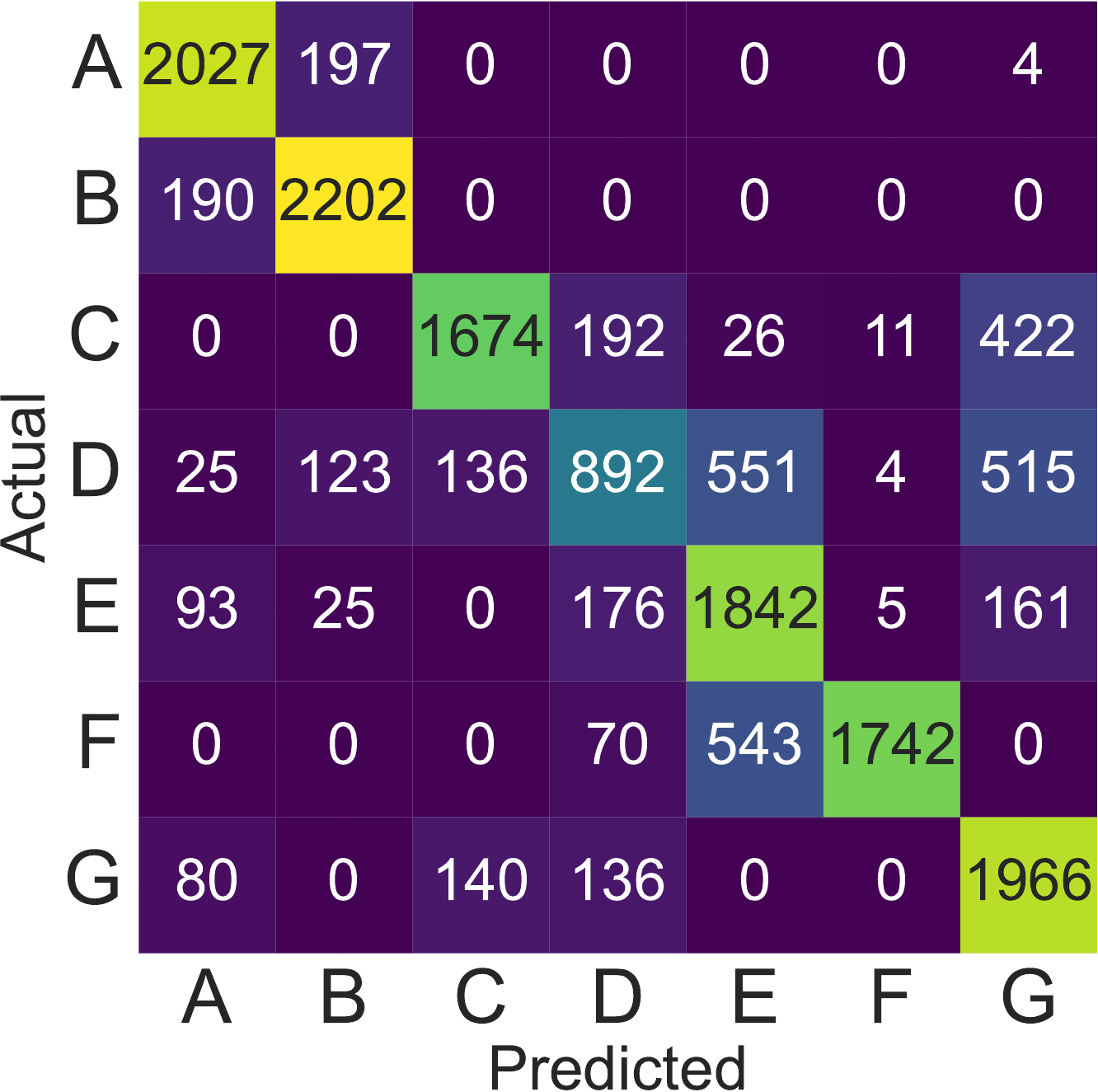}
         \caption{\nofusing{4}}
         \label{fig:cm-no-fused-4}
    \end{subfigure}
    
    \begin{subfigure}[b]{0.455\linewidth}
         \centering
         \includegraphics[width=\textwidth]{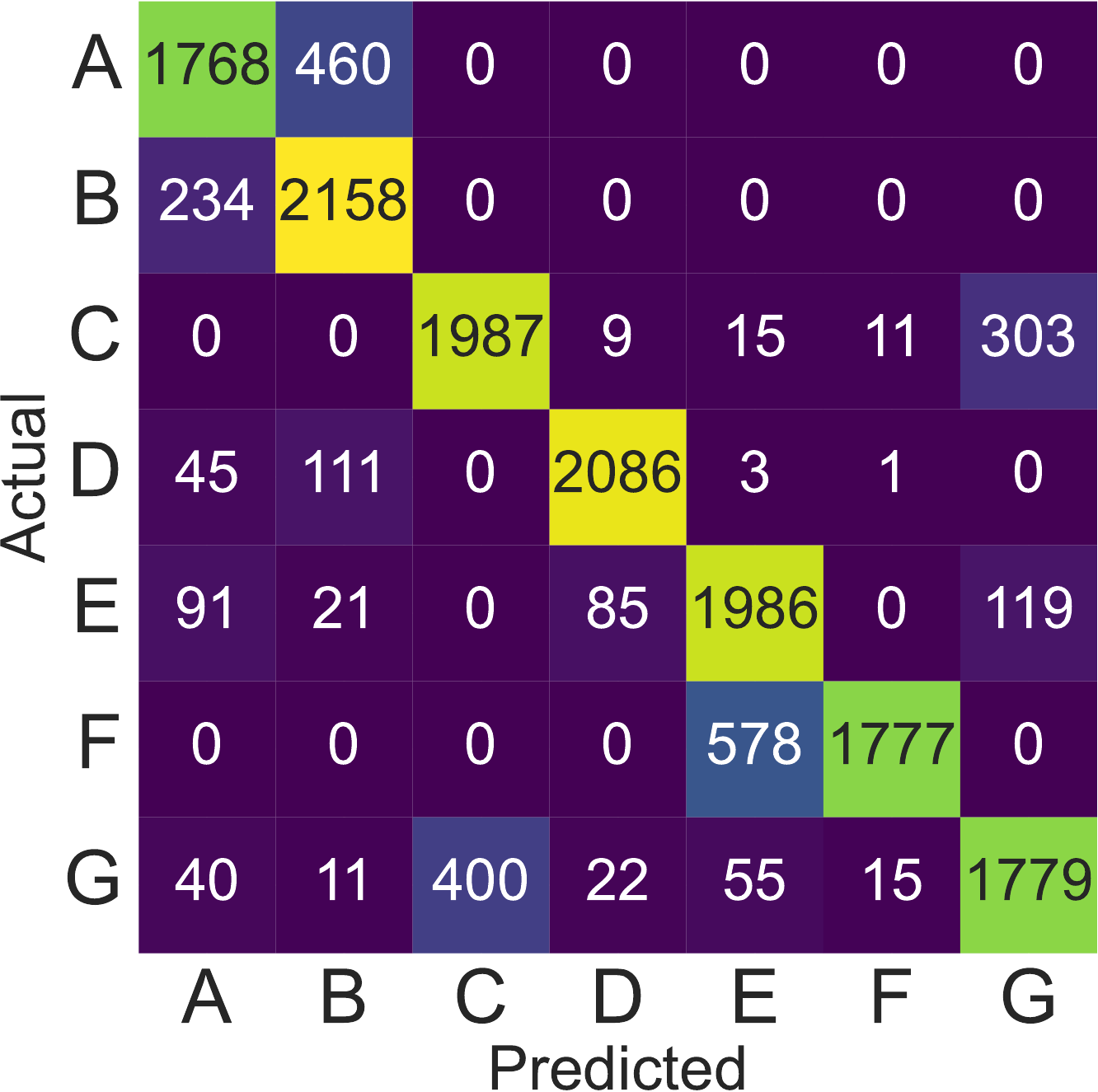}
         \caption{\earlyfusing}
         \label{fig:cm-early-fusing}
     \end{subfigure}
    \hfill
    \begin{subfigure}[b]{0.455\linewidth}
         \centering
         \includegraphics[width=\textwidth]{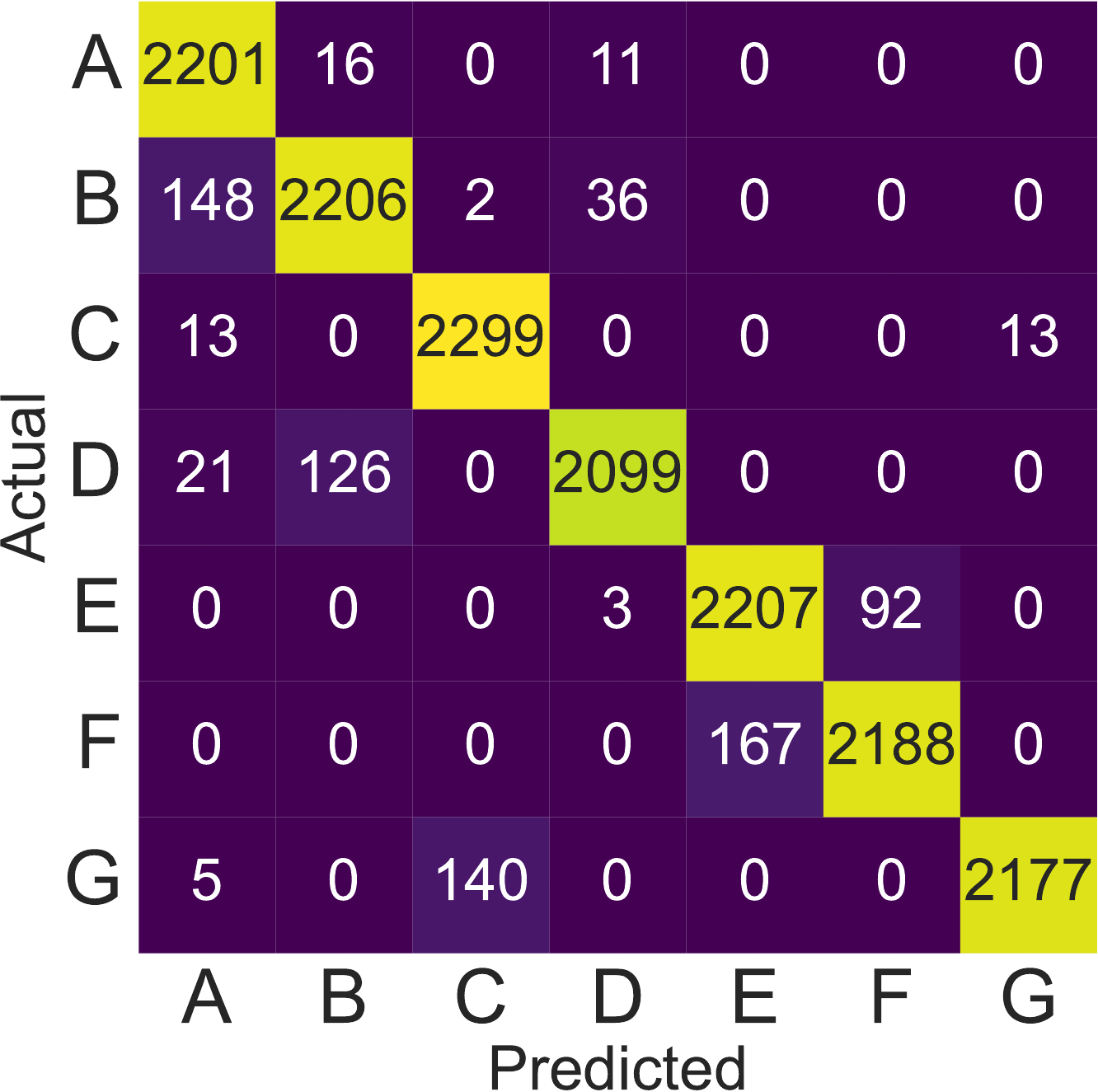}
         \caption{\delayedfusing}
         \label{fig:cm-delayed-fusing}
     \end{subfigure}

    \caption{Confusion matrixes of \deepprobhar for the different fusion strategies. Activities are labelled as: A)~Walk; B)~Run; C)~Squat; D)~Jump; E)~Wave; F)~Clap; G)~Wipe.}
    \label{fig:CM-main-results}
\end{figure}

Finally, we observe that some \glspl{mlp} (\eg\ {\bf MLP 4}) reach $1.00$ accuracy. From \Cref{fig:decision_tree_video}, such a \gls{mlp} is responsible for distinguishing between {\it squatting} and {\it jumping}.
A quick inspection of the confusion matrixes (\Cref{fig:CM-main-results}) reveals that such accuracy is the product of perfect split among those two classes over the test set, indicating that it can be relatively easy for an \gls{mlp} to separate the remaining two activities.

\section{Conclusion}
\label{sec:conclusion}
We introduced  \deepprobhar, a novel neuro-symbolic fusion approach to \gls{har}, and provided initial evidence that simple movements, such as leg or arm movements, which are integral to human activities like running or walking, can be discerned using \wifi signals.
Leveraging declarative knowledge of several activities extracted from a video feed, \deepprobhar achieves results comparable to the state-of-the-art in \gls{csi}-based \gls{har}.
Moreover, as a by-product of the learning process, \deepprobhar generates specialised classifiers for simple movements whose accuracy is on par with that of models trained on finely labelled datasets with a much higher cost.
However, we expect that as the complexity of the events to be detected increases, the neuro-symbolic approach can even outperform only-neural techniques~\cite{han2024empirical}.

In future work, we will examine the efficacy of discerning simple movements when categorising unseen activities, \eg \textit{parkour}.
We shall also evaluate whether the inductive bias provided by the symbolic part enables learning with smaller training dataset sizes w.r.t. other state-of-the-art \gls{har} models.
Second, we intend to utilise the declarative knowledge of \deepprobhar to explain the latent space of the \glspl{vae} employed as input.
This will help us better comprehend the underlying structure and distribution of the data, potentially leading to more precise and efficient models.
Third, we will consider an \gls{edl} \cite{Sensoy2018,sensoy_UncertaintyAwareDeepClassifiers_20} approach to enhance robustness against out-of-distribution data, thereby improving the generalisability and reliability of our models, also across different indoor locations.

\section*{Acknowledgments}
This work was partially supported by the European Office of Aerospace Research \& Development (EOARD) under award number FA8655-22-1-7017 and by the US DEVCOM Army Research Laboratory (ARL) under Cooperative Agreements \#W911NF2220243 and \#W911NF1720196. Any opinions, findings, and conclusions or recommendations expressed in this material are those of the author(s) and do not necessarily reflect the views of the United States government.

This work was conducted while M. Cominelli was affiliated with the DII, University of Brescia, Italy.

The authors thank Dr. Marc Roig Vilamala for his help in debugging part of the DeepProbLog code used in this work.

While preparing this work, the authors used GPT-3.5 and 4.0 to improve readability and language. After using them, the authors reviewed and edited the content as needed, and they take full responsibility for the publication's content.

\bibliographystyle{IEEEtran} 
\bibliography{additional, references, biblio}

\end{document}